\documentclass[aps,prd,reprint,superscriptaddress,showpacs,showkeys,amsmath,amssymb,floatfix,nofootinbib]{revtex4-1}

\usepackage{graphicx}
\usepackage{dcolumn}
\usepackage{bm}
\usepackage{color}
\usepackage[mathlines]{lineno}
\usepackage[none]{hyphenat}
\usepackage{array}
\usepackage{multirow}
\usepackage{booktabs}
\usepackage{siunitx}
\usepackage{cleveref}
\newcolumntype{M}[1]{>{\centering\arraybackslash}m{#1}}

\usepackage[normalem]{ulem}

\begin{document}

\preprint{APS/123-QED}
\title{Signal Yields of keV Electronic Recoils and Their Discrimination from Nuclear Recoils in Liquid Xenon}




\newcommand{\bologna}{\affiliation{Department of Physics and Astrophysics, University of Bologna and INFN-Bologna, 40126 Bologna, Italy}}
\newcommand{\chicago}{\affiliation{Department of Physics \& Kavli Institute of Cosmological Physics, University of Chicago, Chicago, IL 60637, USA}}
\newcommand{\coimbra}{\affiliation{LIBPhys, Department of Physics, University of Coimbra, 3004-516 Coimbra, Portugal}}
\newcommand{\columbia}{\affiliation{Physics Department, Columbia University, New York, NY 10027, USA}}
\newcommand{\lngs}{\affiliation{INFN-Laboratori Nazionali del Gran Sasso and Gran Sasso Science Institute, 67100 L'Aquila, Italy}}
\newcommand{\freiburg}{\affiliation{Physikalisches Institut, Universit\"at Freiburg, 79104 Freiburg, Germany}}
\newcommand{\mainz}{\affiliation{Institut f\"ur Physik \& Exzellenzcluster PRISMA, Johannes Gutenberg-Universit\"at Mainz, 55099 Mainz, Germany}}
\newcommand{\heidelberg}{\affiliation{Max-Planck-Institut f\"ur Kernphysik, 69117 Heidelberg, Germany}}
\newcommand{\munster}{\affiliation{Institut f\"ur Kernphysik, Westf\"alische Wilhelms-Universit\"at M\"unster, 48149 M\"unster, Germany}}
\newcommand{\nikhef}{\affiliation{Nikhef and the University of Amsterdam, Science Park, 1098XG Amsterdam, Netherlands}}
\newcommand{\nyuad}{\affiliation{New York University Abu Dhabi, Abu Dhabi, United Arab Emirates}}
\newcommand{\paris}{\affiliation{LPNHE, Universit\'e Pierre et Marie Curie, Universit\'e Paris Diderot, CNRS/IN2P3, Paris 75252, France}}
\newcommand{\purdue}{\affiliation{Department of Physics and Astronomy, Purdue University, West Lafayette, IN 47907, USA}}
\newcommand{\rpi}{\affiliation{Department of Physics, Applied Physics and Astronomy, Rensselaer Polytechnic Institute, Troy, NY 12180, USA}}
\newcommand{\rice}{\affiliation{Department of Physics and Astronomy, Rice University, Houston, TX 77005, USA}}
\newcommand{\stockholm}{\affiliation{Oskar Klein Centre, Department of Physics, Stockholm University, AlbaNova, Stockholm SE-10691, Sweden}}
\newcommand{\subatech}{\affiliation{SUBATECH, IMT Atlantique, CNRS/IN2P3, Universit\'e de Nantes, Nantes 44307, France}}
\newcommand{\torino}{\affiliation{INFN-Torino and Osservatorio Astrofisico di Torino, 10125 Torino, Italy}}
\newcommand{\ucla}{\affiliation{Physics \& Astronomy Department, University of California, Los Angeles, CA 90095, USA}}
\newcommand{\ucsd}{\affiliation{Department of Physics, University of California, San Diego, CA 92093, USA}}
\newcommand{\wis}{\affiliation{Department of Particle Physics and Astrophysics, Weizmann Institute of Science, Rehovot 7610001, Israel}}
\newcommand{\zurich}{\affiliation{Physik-Institut, University of Zurich, 8057  Zurich, Switzerland}}

\author{E.~Aprile}\columbia
\author{J.~Aalbers}\nikhef
\author{F.~Agostini}\lngs\bologna
\author{M.~Alfonsi}\mainz
\author{F.~D.~Amaro}\coimbra
\author{M.~Anthony}\columbia
\author{F.~Arneodo}\nyuad
\author{P.~Barrow}\zurich
\author{L.~Baudis}\zurich
\author{B.~Bauermeister}\stockholm
% ...
\collaboration{XENON Collaboration}\email{xenon@lngs.infn.it}\noaffiliation

\date{\today}

\begin{abstract}

We report on the response of liquid xenon to low energy electronic recoils below 15\,keV from beta decays of tritium at drift fields of 92\,V/cm, 154\,V/cm and 366\,V/cm using the XENON100 detector. A data-to-simulation fitting method based on Markov Chain Monte Carlo is used to extract the photon yields and recombination fluctuations from the experimental data. The photon yields measured at the two lower fields are in agreement with those from literature; additional measurements at a higher field of 366\,V/cm are presented. 
The electronic and nuclear recoil discrimination as well as its dependence on the drift field and photon detection efficiency are investigated at these low energies. The results provide new measurements in the energy region of interest for dark matter searches using liquid xenon. 

\end{abstract}

\maketitle


\section{\label{sec:intro}Introduction}
The nature of dark matter is one of the most intriguing open physics questions today. According to several theories beyond the Standard Model (e.g., Supersymmetry~\cite{Chattopadhyay:2003xi}), dark matter is comprised of Weakly Interacting Massive Particles (WIMPs) which may interact with atomic nuclei via elastic scattering, resulting in nuclear recoils (NRs). Large detectors that use liquid xenon as a target have played a crucial role in pushing down the sensitivity to dark matter-nucleon scattering cross sections, with the most sensitive results recently reported by the LUX, PandaX, and XENON1T experiments~\cite{Akerib:2016vxi,Tan:2016zwf,Aprile:2017iyp,Cui:2017nnn}. Future large liquid xenon detectors, such as XENONnT~\cite{Aprile:2015uzo}, PandaX-4T~\cite{Liu:2017nphy}, LZ~\cite{Mount:2017qzi} and DARWIN~\cite{Aalbers:2016jon}  will further improve the sensitivity by one to two orders of magnitude.

The dominant background component for these large liquid xenon detectors comes from electronic recoils (ERs). 
A precise modeling of the ER background will reduce the uncertainties of the sensitivity for WIMP elastic scattering searches.
In addition, other dark matter candidates, such as axions~\cite{Sikivie:1983ip,Redondo:2013wwa}, can interact with electrons, resulting in electronic recoil signals~\cite{Aprile:2015ade}. Thus understanding the response of electronic recoils in liquid xenon is also crucial to interpret signals resulting from dark matter-electron interactions. While the response of liquid xenon to low-energy nuclear recoils has been extensively measured~\cite{Manzur:2009hp,Plante:2011hw,Aprile:2013teh,Akerib:2016mzi} with a sufficiently accurate description by the NEST v1.0 model~\cite{Lenardo:2014cva}, the response to  electronic recoils below 10~keV still has large uncertainties~\cite{Aprile:2012an,Baudis:2013cca, Akimov:2014cha,Lin:2015jta, Goetzke:2016lfg,Akerib:2016qlr}, mainly due to the lack of calibration data with adequate statistics in the low-energy region.

Measurements of signal responses to ERs below 15\,keV under three different fields in XENON100 are presented in this paper.
This paper is organized as follows: Section~\ref{sec:datasel} discusses the data taking and event selections. In Section~\ref{sec:detcalib}, we describe the detector calibration for electronic recoils using several monoenergetic sources and a model of the anti-correlation between ionization and scintillation. Section~\ref{sec:sim} details simulations using the empirical microphysics model as in NEST~\cite{NEST:v0.98}. We describe the Bayesian fitting method based on a Markov Chain Monte Carlo (MCMC) technique in section~\ref{sec:fitting_method}. We interpret our results in terms of electron-ion recombination in Section~\ref{sec:recom}, and report the observed ER/NR discrimination for drift fields between 92 and 366\,V/cm in Section~\ref{sec:disc}. Section~\ref{sec:conclusion} summarizes our results.  


\section{\label{sec:exp}Data and Analyses}

\subsection{Data taking and selection}
\label{sec:datasel}
The XENON100 detector was operational from 2009 to 2016 at the Gran Sasso National Underground Laboratory. It collected a total of 477 live-days (48 kg$\cdot$yr) of dark matter data~\cite{Aprile:2016swn}.
The details of the experimental apparatus can be found in~\cite{Aprile:2011dd}. 
The last phase of the XENON100 operation was devoted to a series of calibration campaigns using internal sources such as $^{83\mathrm{m}}$Kr, $^{220}$Rn~\cite{Aprile:2016pmc} and the tritiated methane (CH$_3$T) described here. The tests of these new calibration sources provide guidance for the calibration of the larger XENON1T detector for which external calibration sources are not able to probe the inner part of the target.

The tritiated methane source used in this study was obtained from American Radiolabeled Chemicals, Inc.
A 37\,MBq source was diluted by volumetric expansion into isolated pipettes with 10\,Bq activity.
These pipettes were connected to the XENON100 gas circulation system where they could be individually injected into the system.

\begin{figure*}[!t]
\centering
\includegraphics[width=1.8\columnwidth]{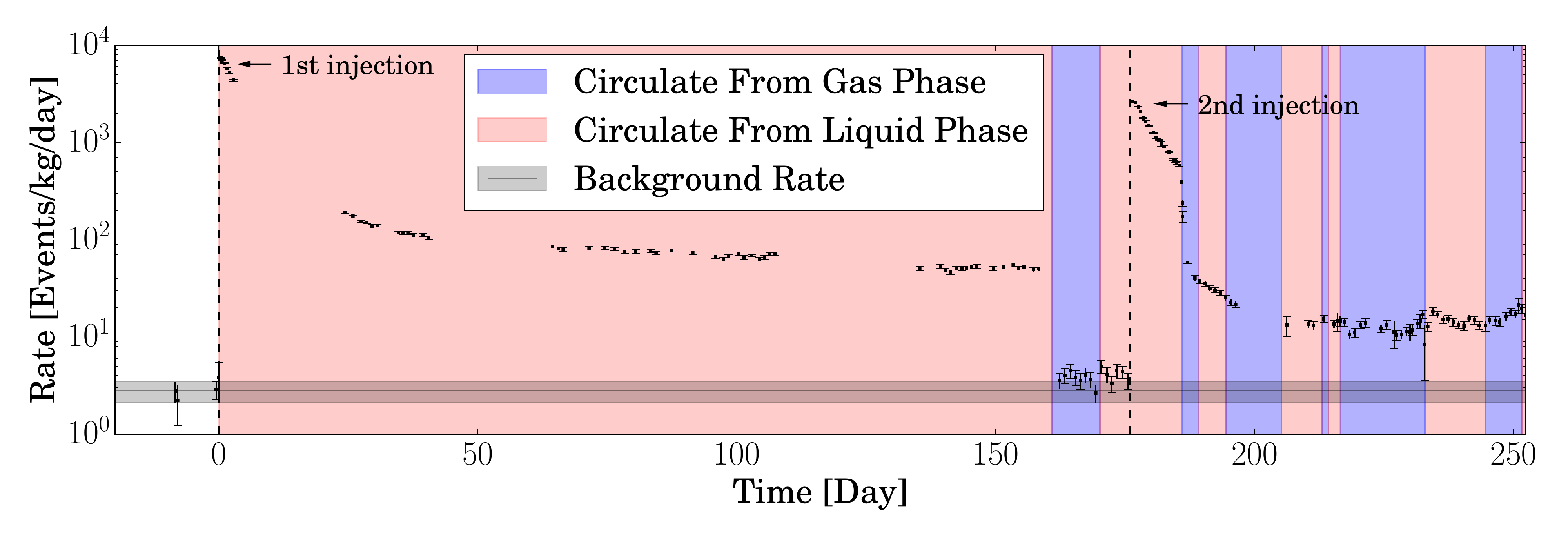}
\caption{Tritiated methane rate starting from the first injection. The left and right vertical black dashed lines represent the first and second tritium injection separately. The periods of circulating the xenon from the bulk liquid and from the gas phase are marked by the red and blue regions, respectively. The horizontal black solid line shows the background rate before tritium injection with the uncertainty in the rate, within 1$\sigma$, given by the black shaded region.
}
\label{fig:rate_data}
\end{figure*}

Following the first source injection in November 2015, the initial tritium event rate in the Time Projection Chamber (TPC) was 7390$\pm$90 events/kg/day. Tritium data at 366\,V/cm drift field was taken. The xenon, from the bulk liquid, was constantly circulated at a speed of about 5 SLPM (standard liter of gas per minute) through a SAES getter purifier where the tritiated methane was removed. The tritium event rate in the detector was reduced, however the tritiated methane removal speed became very slow. Four months after the injection, the tritium event rate was reduced to about 50 events/kg/day. Later, we used an alternative path to circulate the xenon from the gas phase through the purifier. Circulating xenon from the gas phase dramatically improved the removal efficiency and the tritium rate dropped quickly to near zero (1.1$\pm$1.0 events/kg/day after subtracting the background rate). Figure~\ref{fig:rate_data} shows the event rate evolution following different circulation paths. 

The second source injection was performed in May 2016 with an initial tritium event rate of 2640$\pm$20 events/kg/day in the XENON100 TPC. Tritium data was acquired at 92\,V/cm and 154\,V/cm. 
Following that, we circulated gas-phase xenon as well as the xenon from the liquid phase. However, the tritium event rate remained at about 12.8$\pm$1.7 events/kg/day in the end and couldn't be removed further, even with a new xenon purifier. The remaining tritium rate could come from trace contaminants, such as tritiated water or heavy hydrocarbons, which could be removed by a methane purifier as used in \cite{Akerib:2015wdi} but the purifier was not implemented in our setup. 


The tritium event rates during the calibration data taking were at least 3 orders of magnitude higher than the background rate. Neutron calibration data with an $^{241}$AmBe source at the three drift fields were taken as well for the ER/NR discrimination study. 
During the tritium data acquisition, the 662~keV mono-energetic gamma line from an external $^{137}$Cs source was used to monitor the detector conditions, such as the light yield and electron lifetime, which describes the purity of the liquid xenon. The data and detector conditions are summarized in Table~\ref{tab:data}.

\begin{table}[htb]
\small
\centering
\begin{tabular}{
	c
    c
    c
    c
    c
    c@{\,\( \pm \)\,}
    c
    c
    }
\hline\hline
\multirow{2}{*}{Source} & $V_{c}$ & $V_{a}$ & $E_d$  & $t_d^{max}$ & \multicolumn{2}{c}{$\tau_e$} & $N_{FV}$   \\
   & (kV)  & (kV) & (V/cm) & ($\mu$s) & \multicolumn{2}{c}{($\mu$s)} &      ($10^4$) \\
\hline
 & -12  & 4.4 & 366 $\pm$ 24 & 182 & 1470 & 190  &  43.4  \\
CH$_3$T            & -5 & 3.6 & 154 $\pm$ 10 & 202 & 390 & 160  & 11.9\\
           & -3 & 3.6 & 92 $\pm$ 6 & 220 & 590 & 30   & 8.9   \\
\hline       
           & -12 & 4.4 & 366 $\pm$ 24 & 182 & 1490 & 100 &  3.5 \\
 $^{241}$AmBe     & -5 & 3.6 & 154 $\pm$ 10 & 202 & 490 & 130   & 3.6   \\
          & -3 & 3.6 & 92 $\pm$ 6 & 220 & 550 & 60 & 6.5    \\
\hline\hline
\end{tabular}
\caption{Data taking conditions for the ER calibration (CH$_3$T) and NR calibration ($^{241}$AmBe). The voltages on the cathode and anode are $V_c$ and $V_a$, respectively. The volume-averaged drift field, $E_d$, was determined by a 2D finite-element simulation in COMSOL. The uncertainty of $E_d$ represents the change of field strengths from fiducial volume (FV) \#1 to \#7, according to Fig.~\ref{fig:fv}. The relative standard deviation in the field strength within each small FV is less than 2\%. The maximum drift time across the entire drift length is $t_d^{max}$, and $\tau_e$ is the average electron lifetime for each run. The total number of single-scatter events after quality cuts is $N_{FV}$, with $S1$ signals between 3 and 100 PE in the seven FVs used for this study.}  
\label{tab:data}
\end{table}%

The total reflection of the primary scintillation light ($S1$) at the liquid-gas interface, the detector geometry, and the quantum and collection efficiencies of the photomultiplier tubes (PMTs) lead to a non-uniform light collection across the active volume. These factors affect the photon detection efficiency ($PDE$) which is the probability of a scintillation photon being detected by the PMTs. The $PDE$ is a fundamental detector parameter as it influences the energy resolution and threshold of the instrument. To investigate the effect of the $PDE$ on ER/NR discrimination, we chose data in small fiducial volumes (FVs) within which approximately uniform $PDE$s can be obtained. 

\begin{figure}[htbp]
\begin{center}
\includegraphics[width=1.0\columnwidth]{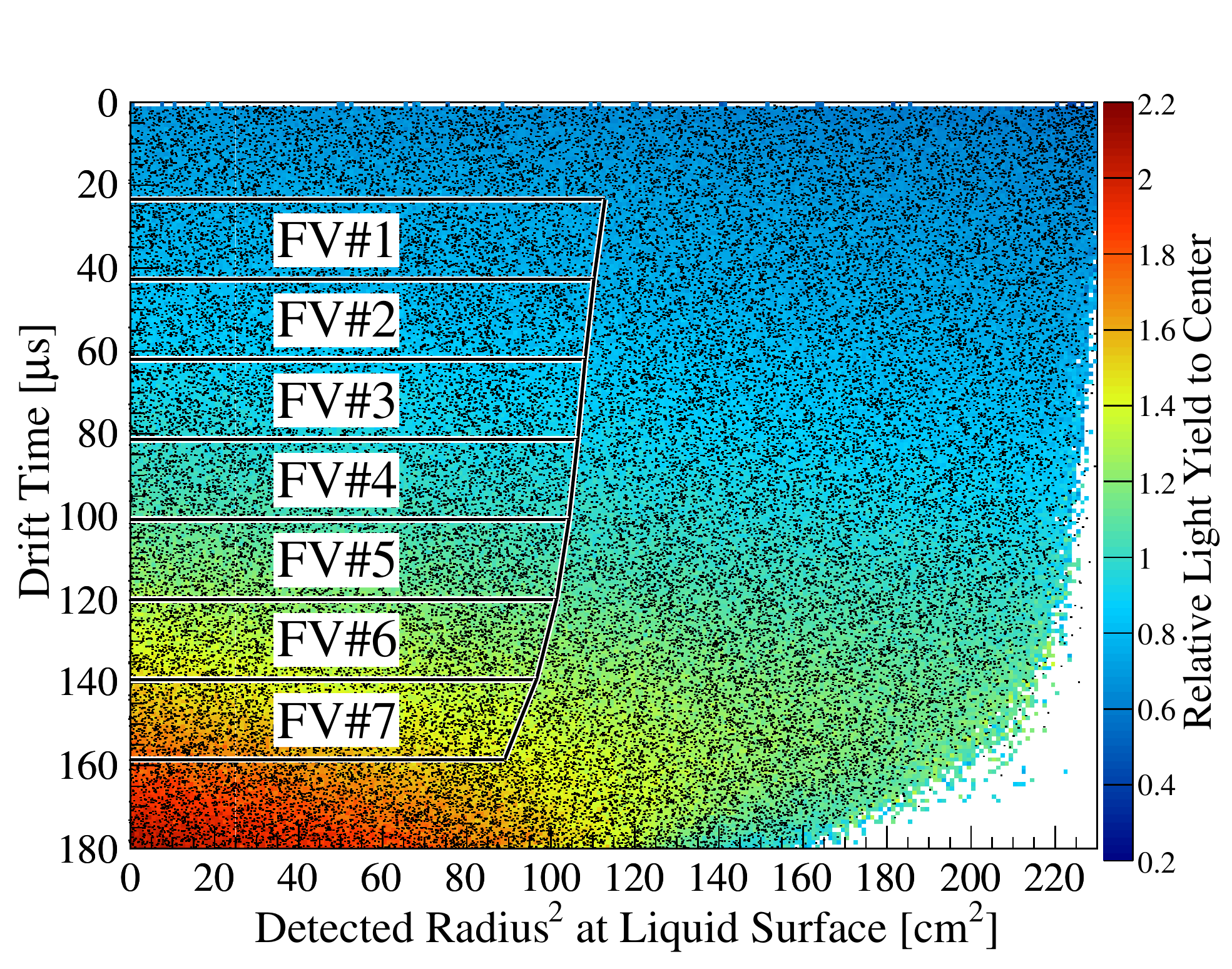}
\caption{The uniform spatial distribution of  tritium beta-decay events (black dots), under the drift field of 366\,V/cm, overlaid with the $S1$ light yield (photoelectrons per keV energy deposition) relative to that of the center (FV\#4) from the $^{83\mathrm{m}}$Kr (41.5~keV) calibration. The radial positions used in this analysis are not corrected for the field distortion at the bottom corner of the TPC~\cite{Aprile:2011dd}, thus the event locations shown in this plot are those detected at the liquid surface. To avoid the systematic effects due to non-uniform drift field, we chose 50\% of events from the FV in the central part of the TPC for the following analysis. The central volume, after removing the top and bottom parts, is further divided into seven small FVs, each with a different $S1$ photon detection efficiency which increases from the top to the bottom. }
\label{fig:fv}
\end{center}
\end{figure}

The FVs are defined by choosing, in radial positions, 50\% of the  tritium events, which are expected to be distributed uniformly in the volume quickly after the injection. Due to distortions of the electric fields, the event radial positions detected at the liquid surface are shifted to the inner volume, especially for the events with large drift time, resulting in a curled edge for the FVs. We further divided the selected volume containing 50\% of the events into nine small slices, equally spaced in drift time. The top and bottom slices are not used in the study to avoid systematic effects due to drift field distortion near the edge and the surface of the detector. The seven small FVs used in the study, each corresponding to a liquid xenon mass of about 4.0~kg, are shown in Fig.~\ref{fig:fv}. The small FVs minimize the position-dependent $S1$ and $S2$ signal variations, reaching less than 6\% for $S1$ and 5\% for $S2$. 
These signal variations are caused by the spatial dependence in the detection efficiency, which is accounted for in the simulations illustrated in Eqs.~\ref{eq:PhotonDetection} and ~\ref{eq:ElectronDetection}.
We have collected more than $10^4$ ER events in each of the small FVs to have sufficient statistics to probe the ER rejection power.

\subsection{Detector Calibration}
\label{sec:detcalib}

\begin{figure}[htbp]
\begin{center}
\includegraphics[width=1\columnwidth]{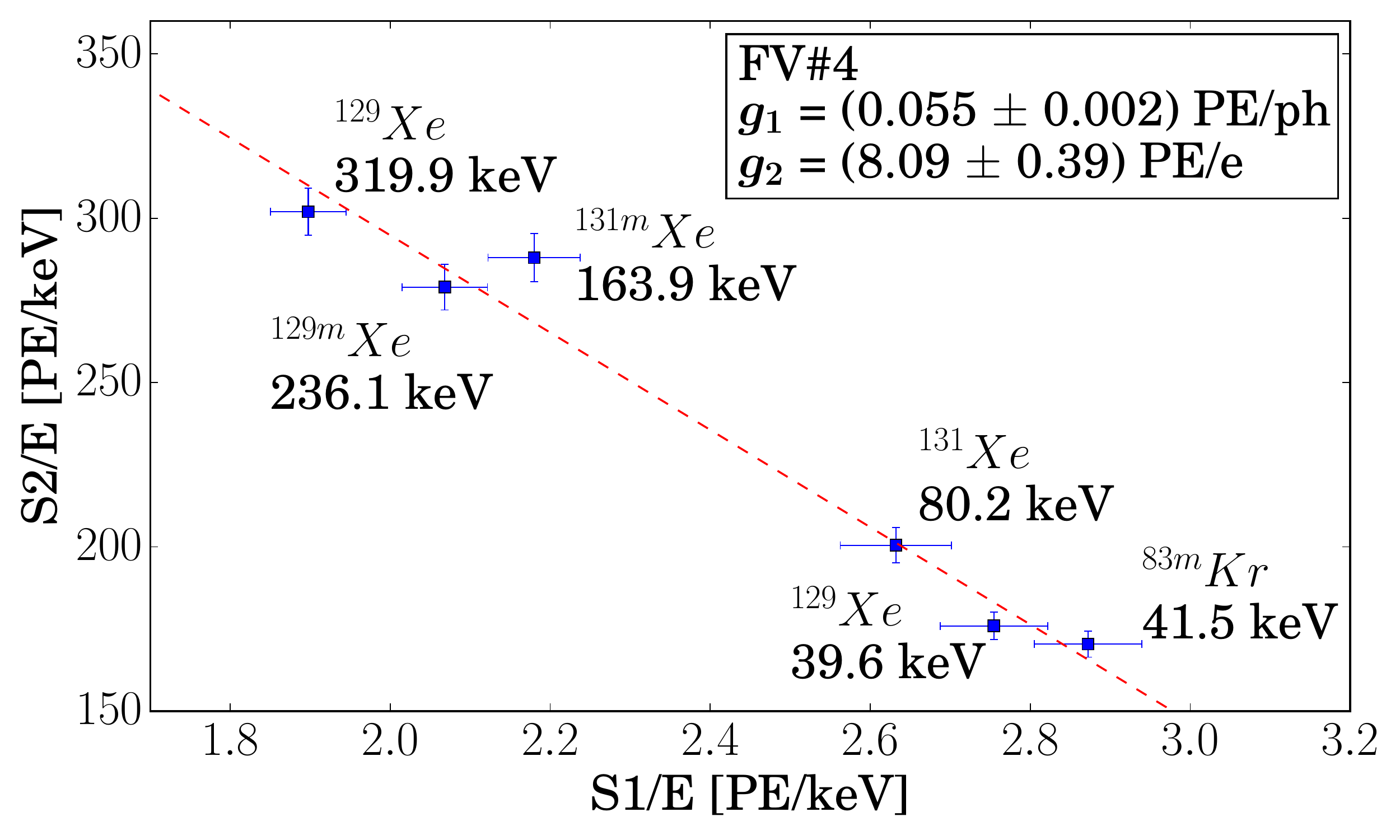}
\caption{Anti-correlation between $S1$ and $S2$ signals for events at different energies. A linear anti-correlation fit is applied to obtain the $g_1$ and $g_2$ values. This plot is for the central fiducial volume (FV\#4) at 366\,V/cm. The energy spectrum of the NR component and the 39.6 and 80.2 keV gamma lines from inelastic scattering are obtained from the Geant4 simulation and converted to photons and electrons separately, according to the NR and ER models in NEST. We then subtracted the average charge/light yields from the NR component to obtain the correct yields for the two pure gamma lines. The 41.5\,keV events from $^{83\mathrm{m}}$Kr are from the combination of two transitions (32.1\,keV and 9.4\,keV)~\cite{Baudis:2010Kr83m}. 
}
\label{fig:g1g2}
\end{center}
\end{figure}

The expectation values of photon and electron gains, $g_1=\langle S1 \rangle/n_{ph}$ and $g_2=\langle S2 \rangle/n_e$, defined as the fractions of detected photoelectrons of S1 and S2 signals to the number of emitted photons and electrons, are key parameters for the detector characterization. The mono-energetic lines used in the $g_1$ and $g_2$ calibration are the 39.6\,keV from $^{83m}$Kr and the activated xenon lines during or after the $^{241}$AmBe neutron calibration. The $g_1$ and $g_2$ values in each small fiducial volume under each scanned field are obtained by applying a linear anti-correlation fit, according to Eq.1, on these energy points with an average energy W to produce a quantum (photon or electron) fixed at $13.7\pm0.2$~eV~\cite{Dahl:2009nta}. We show in Fig.~\ref{fig:g1g2} an example fit for FV\#4. 

\begin{equation}
\begin{split}
\label{ER}
\frac{E}{W} = n_{ph} + n_e = \frac{S1}{g_1} + \frac{S2}{g_2}.
\end{split}
\end{equation}

\begin{figure}[htbp]
\begin{center}
\includegraphics[width=1\columnwidth]{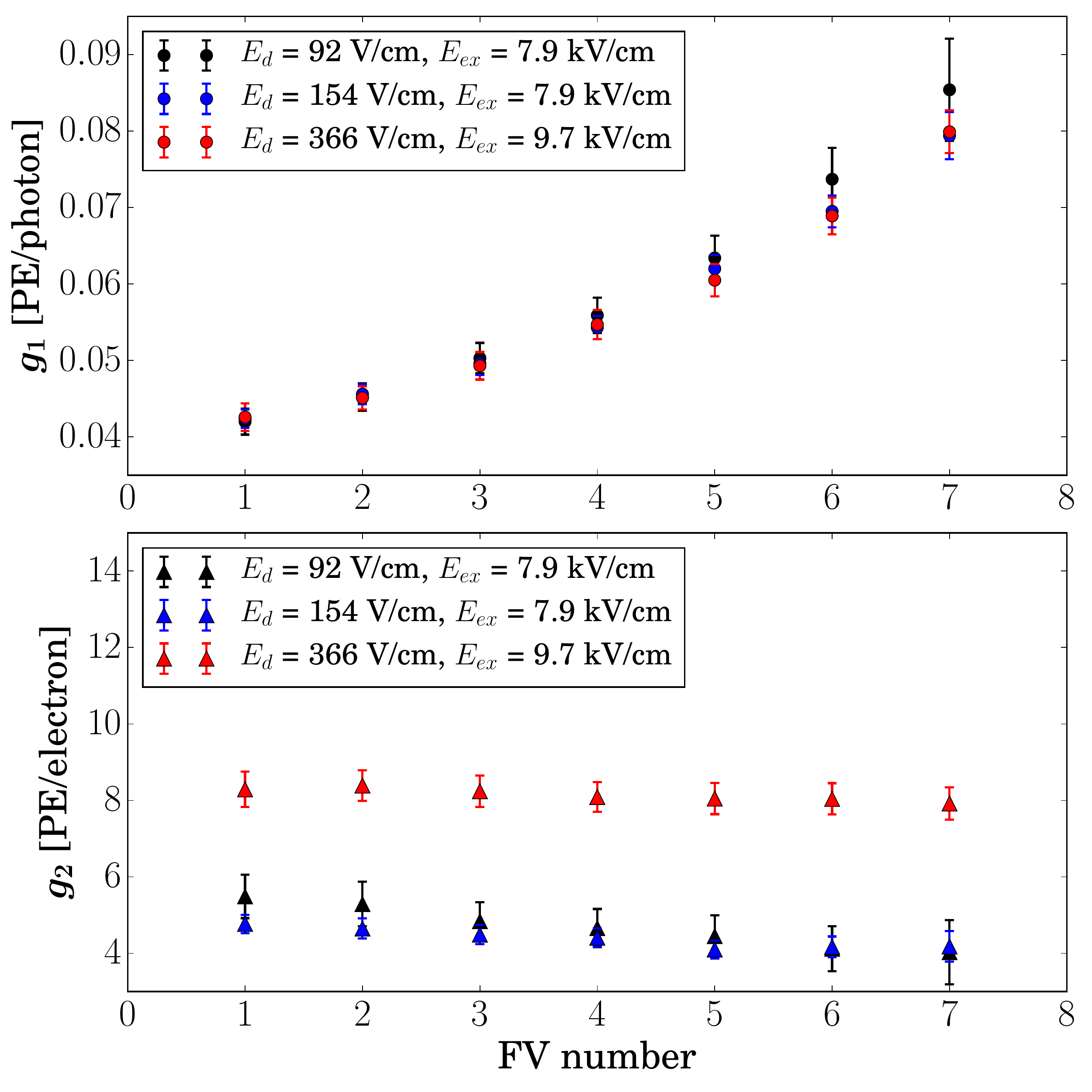}
\caption{The $g_1$ and $g_2$ values for seven FVs (see FV numbering in Fig.~\ref{fig:fv}) at three different field configurations. The $g_1$ values depend only on the detector geometry and PMT quantum/collection efficiency, and they are consistent at the three different fields. The $g_1$ values increase towards the lower part of the detector. The $g_2$ values depend on the liquid purity (electron lifetime) and also on the electron extraction efficiency, thus show lower values at a lower extraction field. The $E_{ex}$ values are obtained based on a 2D finite element simulation in COMSOL, with an uncertainty of 0.1~kV/cm due to the uncertainty of the liquid level. Using a parallel plate approximation will result in extraction fields 0.3 kV/cm higher than those from simulation. The electron extraction efficiency across the liquid-gas interface, calculated based on the ratio between $g_2$ and the single electron gas gain, is 96$\pm$2\% for a $V_a$ of 4.4~kV ($E_{ex}$=9.7~kV/cm), and is about 84$\pm$6\% at 3.6~kV ($E_{ex}$=7.9~kV/cm). 
}
\label{fig:g1g2all}
\end{center}
\end{figure}

 The $g_1$ and $g_2$ values obtained with this method for other FVs and at other fields are shown in Fig.~\ref{fig:g1g2all}. In this study, we performed the analysis in each small FV where the spatial variations of $S1$ and $S2$ signals are rather small(6\% for $S1$ and 5\% for $S2$), thus the $S1$ and $S2$ signals are not corrected for position dependence.

\subsection{Signal Simulation}
\label{sec:sim}

Simulations of the signal responses to the tritiated methane source are performed.
They take into account both the microphysics of the signal production in liquid xenon and the detection, amplification and reconstruction of the signals by the XENON100 detector and software.

The empirical microphysics model introduced by NEST~\cite{NEST:v0.98} is used in the simulation except for the parameterization of the recombination.
The model used in this work describes the production of photons $n_{ph}$ and electrons $n_e$ following an energy deposition $E$ in liquid xenon.

The total number of quanta, $n_q=n_{ph}+n_e$, has the intrinsic fluctuation $n_q \sim $N$(E/W, \sqrt{FE/W})$ due to the Fano process~\cite{Fano_Process}, where N represents the normal distribution and $F$=0.059 is the Fano factor from Doke's estimation~\cite{Doke_Fano_Estimate}.

The signal production process consists of several steps. First, excitons and electron-ion pairs are produced following the energy deposition. Excitons, $n_{ex} \sim \textrm{Binom}(n_q, \alpha/(1+\alpha))$, directly decay and emit light. Here $\alpha=\langle n_{ex}/n_i \rangle$ is the mean number of excitons ($n_{ex}$) to ions ($n_i=n_q - n_{ex}$) ratio and has a value between 0.06-0.20~\cite{NEST:v0.98} for electronic recoils.

Second, a fraction of electron-ion pairs recombine, with a  recombination fraction $r$, to form excitons and subsequently decay to produce additional scintillation photons. The recombination fraction $r$ depends on the energy and field present in the liquid, and it has a non-negligible intrinsic fluctuation $\Delta r$~\cite{Akerib:2015wdi}.  We assume a truncated-Gaussian distributed recombination fluctuation with $r$ in interval of (0,1) in this work.

\begin{equation}
\label{eq:RecombinationAndFluctuations1}
r \sim\textrm{N}\left( \langle r\rangle, \Delta r, 0, 1 \right).
\end{equation}

Finally, the total number of photons ($n_{ph}$) and electrons ($n_e$) produced after the entire process can be written as:

\begin{equation}
\label{eq:RecombinationAndFlucuations}
\begin{split}
n_{ph} - n_{ex} \sim & \textrm{Binom}(n_i, r), \\
n_e + n_{ph} = & n_q.
\end{split}
\end{equation}

\begin{figure}[htpb]
\includegraphics[width=1.0\columnwidth]{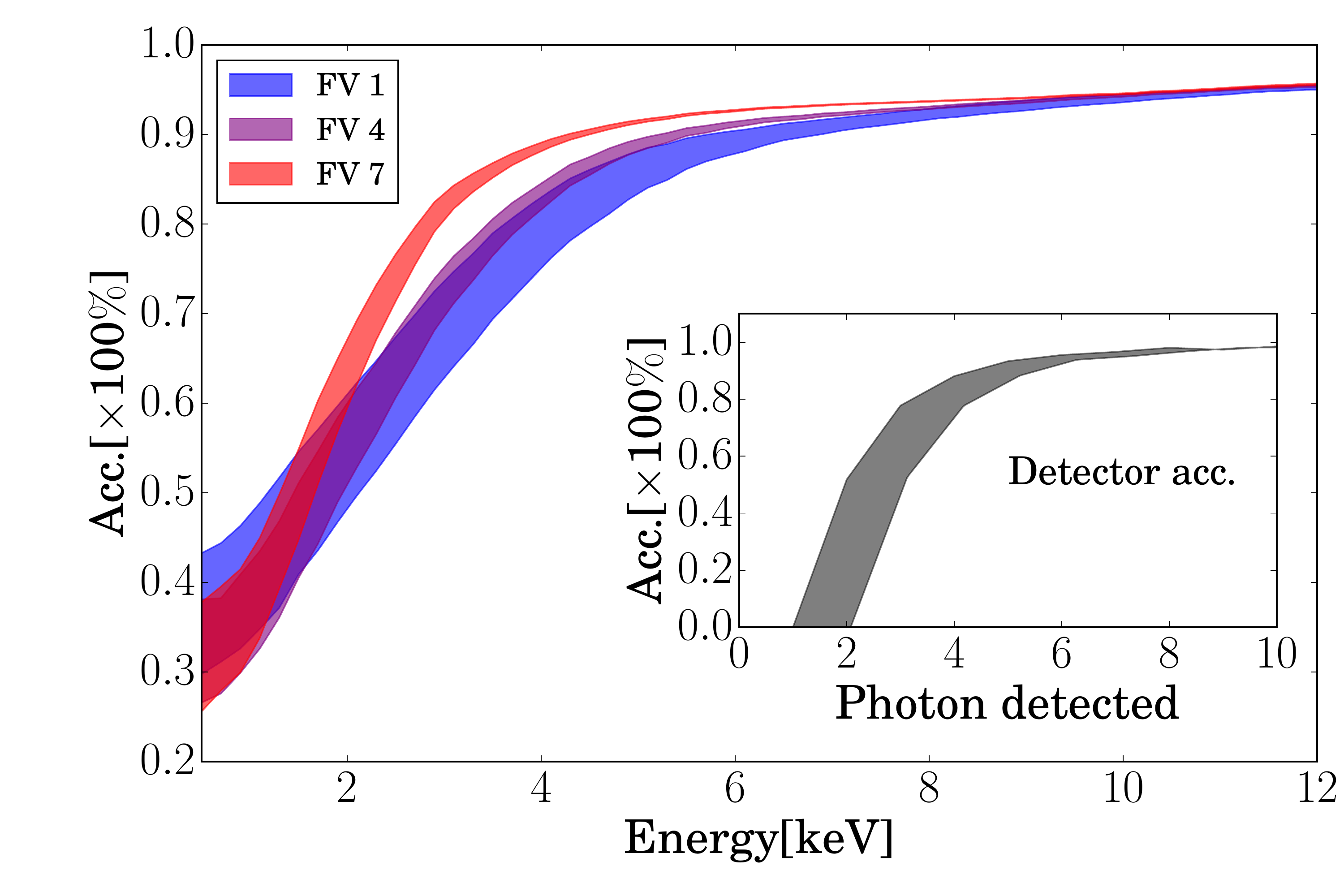}
\caption{
The signal detection efficiencies in this study for sub-volume 1, 4, 7 as a function of recoil energy are shown in blue, purple and red, respectively.
The inset shows the signal detection efficiency from the S1 coincidence requirement (black), which is the dominant contribution to the overall detection efficiency and is a function of detected photon number.
The shaded regions represent the 15.4\%-84.6\% credible region.
}
\label{fig:Efficiency}
\end{figure}

\begin{figure*}[htbp]
\begin{center}
\includegraphics[width=1.8\columnwidth]{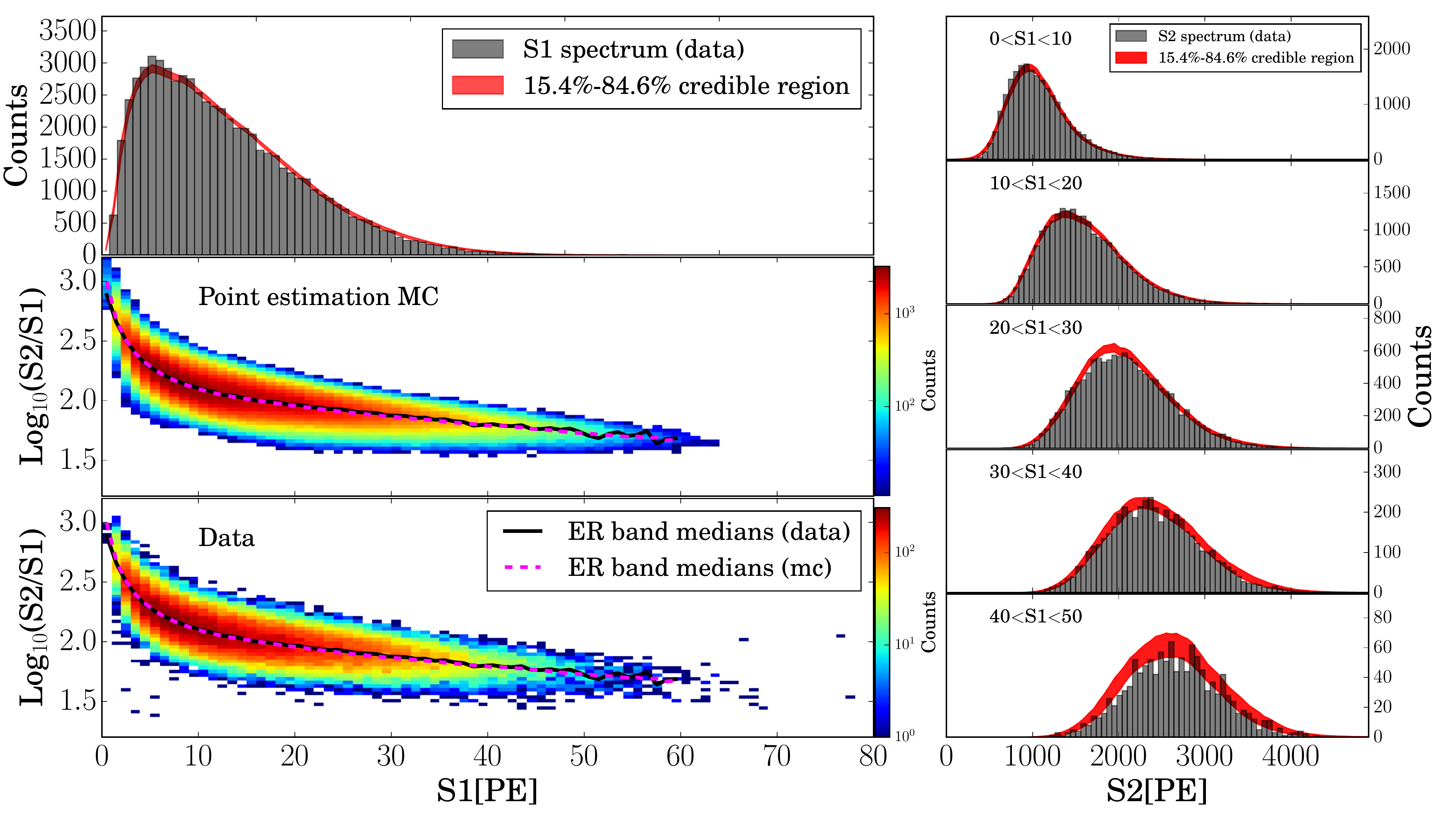}
\caption{
The comparison between the MC fit result (left-middle) and the data (left-bottom), in the form of the 2-D distribution in Log$_{10}(S2/S1)$ versus $S1$ parameter space for FV\#4 at 366~V/cm field. $S1$ spectra (top-left panel) and $S2$ spectra at different $S1$ slices (right panels) are shown together with the 15.4\%-84.6\% credible region from the fit in red.
A goodness-of-fit test using the method in \cite{GelmanMCMCGoF} is performed upon the S2 spectra matching, which gives a p-value of 0.01, 0.16, 0.10, 0.37, 0.38 for the S2 spectra in S1 slices of 0-10\,PE, 10-20\,PE, 20-30\,PE, 30-40\,PE, 40-50\,PE, respectively.
The low p-value for the S2 spectrum matching in S1 slice of 0-10\,PE is caused by the uncertainty of acceptance modeling.    
}
\label{fig:mc_compare_to_data}
\end{center}
\end{figure*}

The photons ($n_{ph}$) are detected by the PMTs as the prompt scintillation signal ($S1$). Photons reaching the photocathode of each PMT have a probability $p_{dpe}$ to produce double photoelectrons as observed in ~\cite{DPEMeasurement}, such that $g_1 = PDE\cdot (1+p_{dpe})$. The $PDE$ and thus $g_1$ depend on the event position, where this dependence is obtained using monoenergetic calibrations of the detector. The number of detected primary photons $n_{dph}$ and detected photoelectrons $n_{pe}$ for $S1$ can be written as:
\begin{equation}
\label{eq:PhotonDetection}
\begin{split}
PDE  = & g_1/(1+p_{dpe}),\\
n_{dph} \sim & \textrm{Binom}\left( n_{ph}, PDE \right),\\
n_{pe} - n_{dph} \sim & \textrm{Binom} \left( n_{dph}, p_{dpe} \right). \\
\end{split}
\end{equation}

The electrons ($n_e$) are drifted with an efficiency $\epsilon_d$, affected by the losses due to capture by electronegative impurities in the liquid, and then extracted into the TPC gas layer with an efficiency $\epsilon_{ext}$ determined by the extraction field. The electrons are accelerated in a stronger field in the gas phase, producing proportional scintillation photons~\cite{Lansiart:1976}. The number of extracted electrons $n_{ext}$ and the number of detected $S2$ photo-electrons, $n_{prop}$, from proportional scintillation, can be written as:

\begin{equation}
\label{eq:ElectronDetection}
\begin{split}
n_{ext} \sim & \textrm{Binom}\left( n_e, \epsilon_{d} \cdot \epsilon_{ext} \right), \\
n_{prop} \sim & \textrm{N}\left(n_{ext} G, \sqrt{n_{ext}} \Delta G \right),
\end{split}
\end{equation}
where $G$ and $\Delta G$ are the single electron gain and its associated standard deviation.
$G$ is x-y dependent and measured from the single-electron spectrum~\cite{Aprile:2013blg}.
$G$ and $\Delta G$ include the effects of gas amplification, detection efficiency of the proportional scintillation due to geometrical coverage and the PMT responses to proportional light, and the associated fluctuations.
The product of $G$, $\epsilon_d$ and $\epsilon_{ext}$ is the $g_2$ value.

The prompt and proportional scintillation signals are digitized by the XENON100 data acquisition system, and then reconstructed in photoelectrons as $S1$ and $S2$, respectively.
During reconstruction, the signal is slightly biased because of the data compression logic of the digitizer~\cite{Aprile:2011dd}, the PMT resolution, and the effect of noise on the baseline calculation.
The reconstructed $S1$ and $S2$ signals are written as,
\begin{equation}
\label{eq:ReconstructionFluctuation}
\begin{split}
S1/n_{pe} - 1 \sim & \textrm{N}\left(\delta_{s1}, \Delta \delta_{s1}\right), \\
S2/n_{prop} - 1 \sim & \textrm{N}\left(\delta_{s2}, \Delta \delta_{s2}\right),
\end{split}
\end{equation}
The bias after reconstruction is modeled as Gaussians with means $\delta_{s1}$ $\delta_{s2}$ and standard deviations $\Delta \delta_{s1}$ $\Delta \delta_{s2}$. These are estimated by reconstructing simulated waveforms that take into account actual $S1$ and $S2$ pulse shapes along with realistic electronic noise.

The signal detection efficiency in this study is evaluated in similar way as in~\cite{Aprile:2016swn}, except for the S1 coincidence requirement. The efficiency for the S1 coincidence requirement, together with the signal reconstruction efficiency of the software, is estimated using a Monte Carlo waveform simulation which implements the shapes of S1s and S2s, the contamination of noise in waveforms, and the signal reconstruction thresholds.
This efficiency is a function of detected photon number and is the dominant contribution to the overall efficiency, as shown in the inset of Fig.~\ref{fig:Efficiency}. Also shown in Fig.~\ref{fig:Efficiency} are the overall efficiencies as a function of deposited energy for different volumes. The differences in the overall efficiency for different volumes are caused by the different $PDE$s. The volume closer to the bottom part of the detector has a higher $PDE$ and thus better efficiency for detecting low-energy recoils.
The efficiency for detecting very low-energy recoils ($<$2\,keV) is not zero because of the fluctuation of reconstructed S1s, as illustrated in \Crefrange{eq:RecombinationAndFluctuations1}{eq:ReconstructionFluctuation}.



\subsection{\label{sec:fitting_method}Fitting Method and Results}

\begin{figure*}[htpb]
\includegraphics[width=1.5\columnwidth]{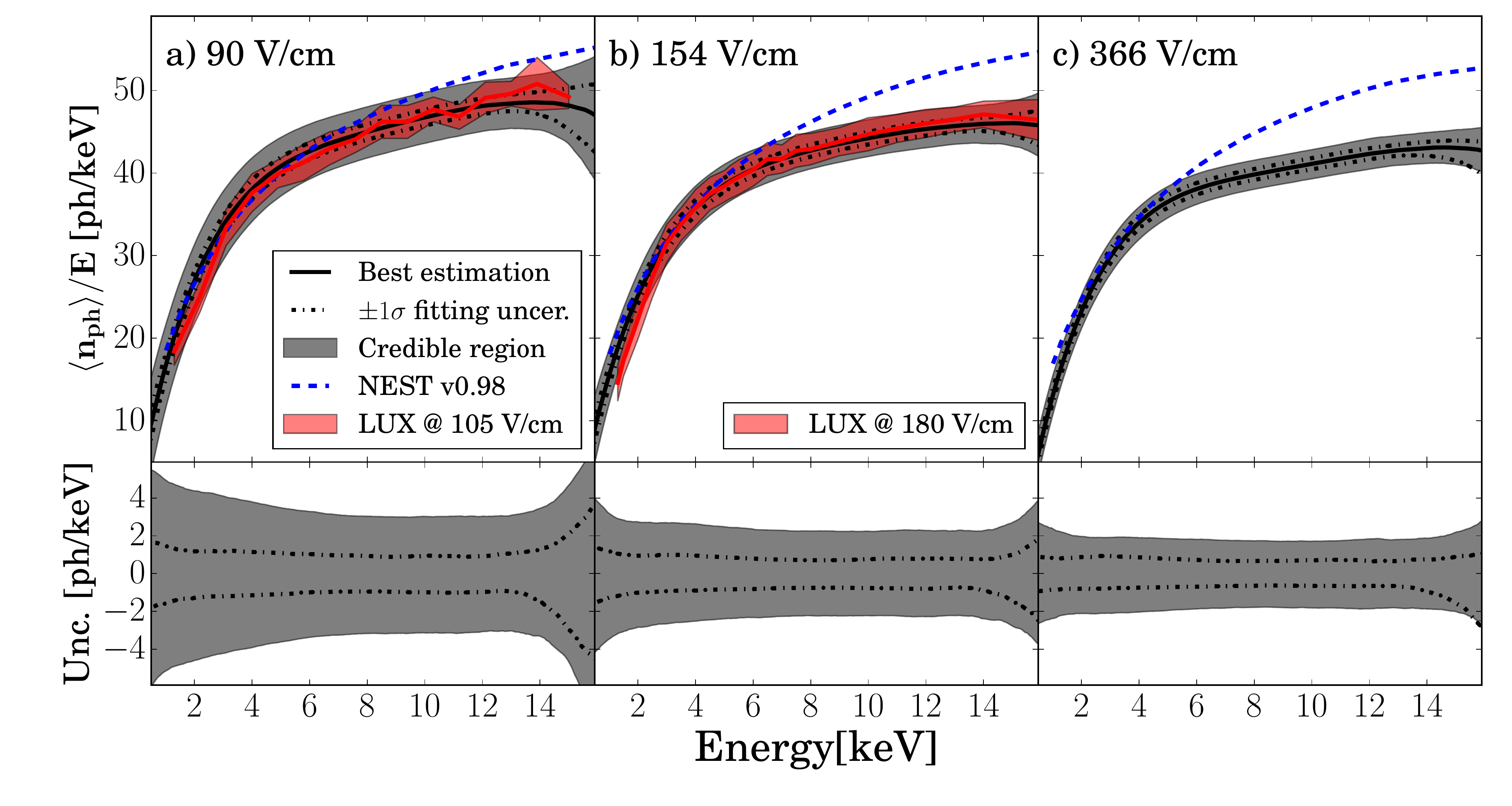}
\includegraphics[width=1.5\columnwidth]{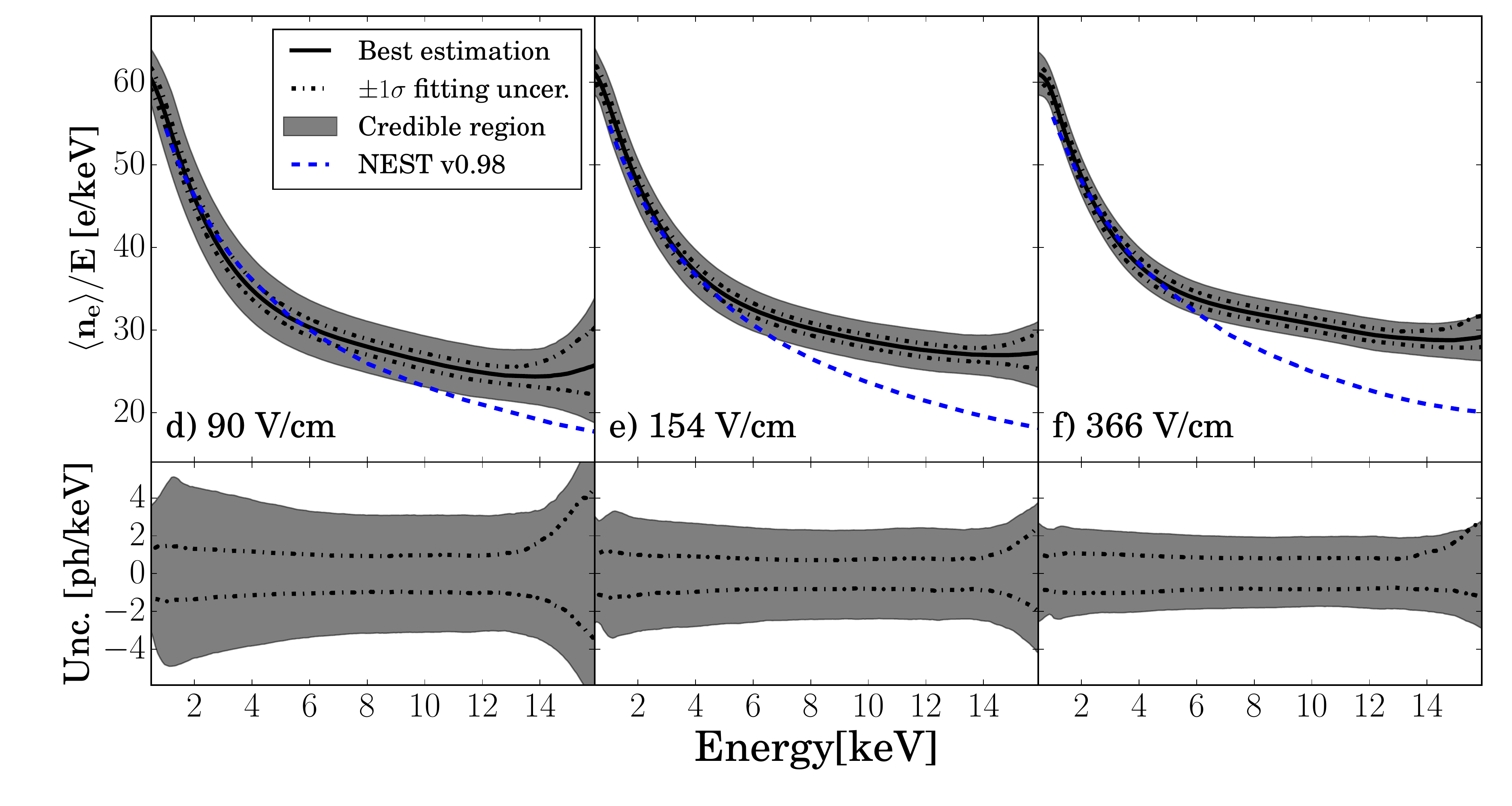}
\includegraphics[width=1.5\columnwidth]{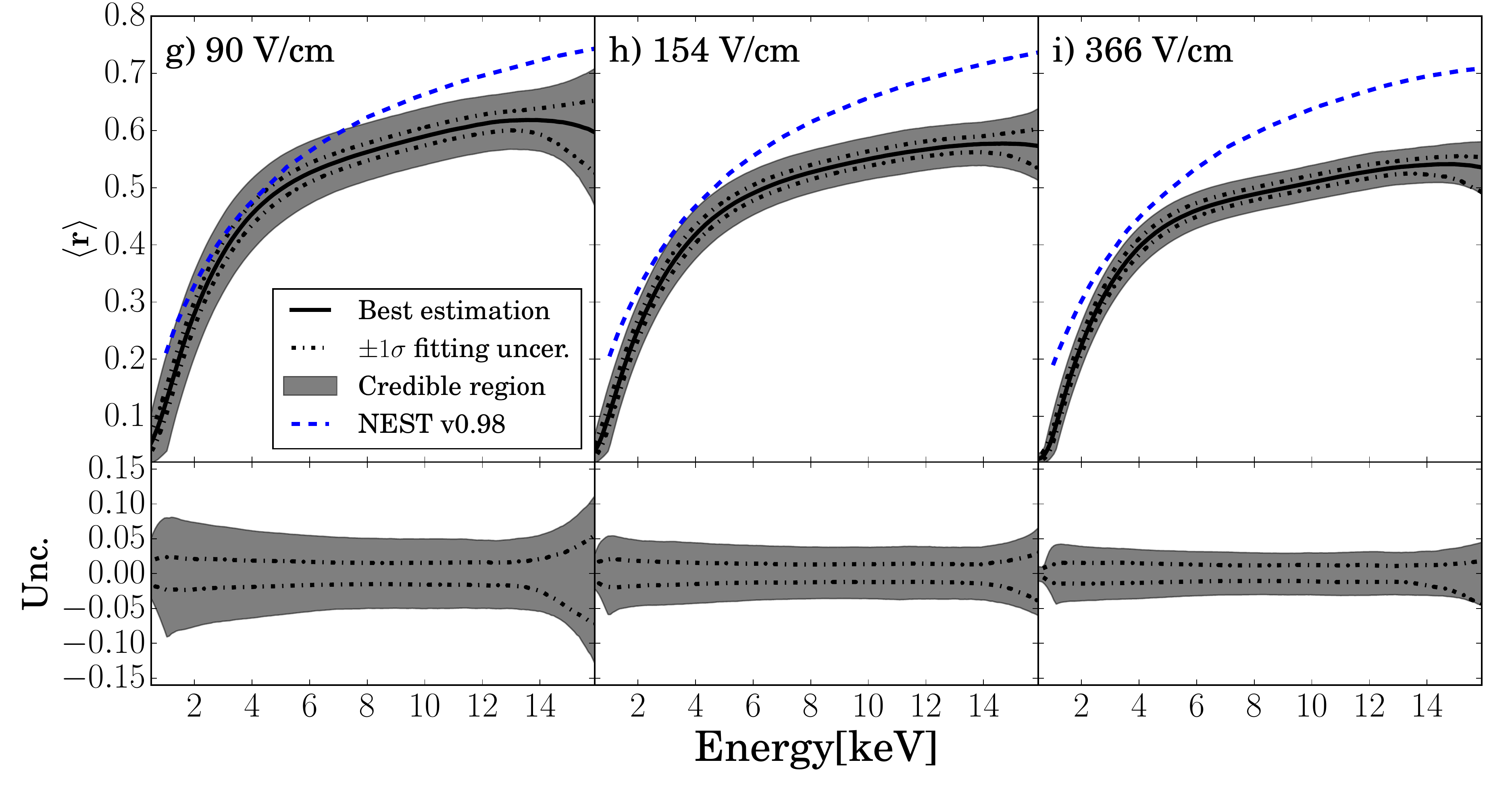}
\caption{
The best estimates for photon yields $\langle n_{ph} \rangle/E$, charge yields $\langle n_e \rangle/E$, and mean recombination fraction $\langle r \rangle$ as a function of deposited energy obtained from the fit are shown in the upper, middle, and lower panels, respectively, for three drift fields. The solid lines represent the mean values and the shaded regions indicate the 15.4\% to 84.6\% credible regions of $\langle n_{ph} \rangle/E$, $\langle n_e \rangle/E$ and $\langle r \rangle$. The dot-dashed lines indicate the fitting uncertainties. Predictions from NEST v0.98~\cite{NEST:v0.98} (dashed blue lines) and measurements from LUX~\cite{Akerib:2015wdi} (red solid lines and shaded regions) are shown for comparison where available.
}
\label{fig:PYandCYandR}
\end{figure*}

A binned Maximum Likelihood Estimation (MLE) analysis of Log$_{10}(S2/S1)$ vs. $S1$ in 2D signal space is performed to extract the electronic recoil signal response model below the 18.6\,keV endpoint of the tritium beta decay.
The likelihood is constructed as:

\begin{equation}
\begin{split}
\label{LikelihoodConstruction}
\mathcal{L} =& \prod_{i,j} \textrm{Poiss} \left( D_{i,j} | fM_{i,j} \right) \times \\
             & \prod_k \textrm{N} \left( \theta_k ; \mu_k, \sigma_k \right) \times \\
             & \prod_n \textrm{Uniform} \left( \phi_n ; \kappa_n, \Theta_n \right),
\end{split}
\end{equation}
where D$_{i,j}$ and M$_{i,j}$ are the counts in each bin from data and simulation, respectively. 
The simulated event rate is scaled by $f = N_{obs} / N_{sim}$ where $N_{obs}$ is the total number of events in the tritium data, with $S1$ in the range of 0 to 80\,PE, and $N_{sim}$ is the total number of simulated events. The nuisance parameters $\theta_k$ are constrained by Gaussian priors with mean $\mu_k$ and standard deviation $\sigma_k$. The nuisance parameters $\phi_n$ are constrained by uniform priors with $\kappa_n$ and $\Theta_n$ being the lower and upper boundaries, respectively.
The nuisance parameters $g_1$, $g_2$, electron lifetime $\tau_e$ and $W$ value are constrained by Gaussian priors. 
Parameters such as exciton-to-ion ratio $n_{ex}/n_i$, double PE emission fraction $p_{dpe}$, event reconstruction efficiency and bias parameters are constrained by uniform priors. 
The constraints for $g_1$ and $g_2$ are shown in Fig.~\ref{fig:g1g2all}, and
$\tau_e$ is listed in Table~\ref{tab:data}.
The constraints for $W$, $n_{ex}/n_i$ and $p_{dpe}$ are taken as ($13.7\pm0.2$)\,eV~\cite{Dahl:2009nta}, 0.06-0.20~\cite{NEST:v0.98} and 0.18-0.24~\cite{DPEMeasurement}, respectively.
The tritium beta decay spectrum is obtained using the calculation in~\cite{BetaSpecKatrin}.

We chose the affine invariant Markov Chain Monte Carlo (MCMC)~\cite{AIMCMC, EMCEE} for maximizing the likelihood and sampling the parameter space. 
The advantages of using MCMC are that it converges relatively quickly given a large number of parameters, and that it can accurately address uncertainties.
The simulation is done at each iteration of MCMC's random walking, which naturally takes care of the uncertainty from the finite statistics in the simulation.
The ratio of the statistics between simulation and data is about 10.
The result comes in the form of Bayesian posteriors, and we will define ``point estimation'' as the posterior median in the rest of the paper.
The comparison of the fit result to data is shown in Fig.~\ref{fig:mc_compare_to_data}. 
The background event rate is four orders of magnitude lower than the event rate from the tritium beta decays in the ER band, thus has negligible impact on the fitting results and the ER leakage fraction studies in Sec~\ref{sec:disc}.


\section{\label{sec:recom} Recombination Factor and Fluctuation}

\begin{figure*}[htpb]
\includegraphics[width=1.5\columnwidth]{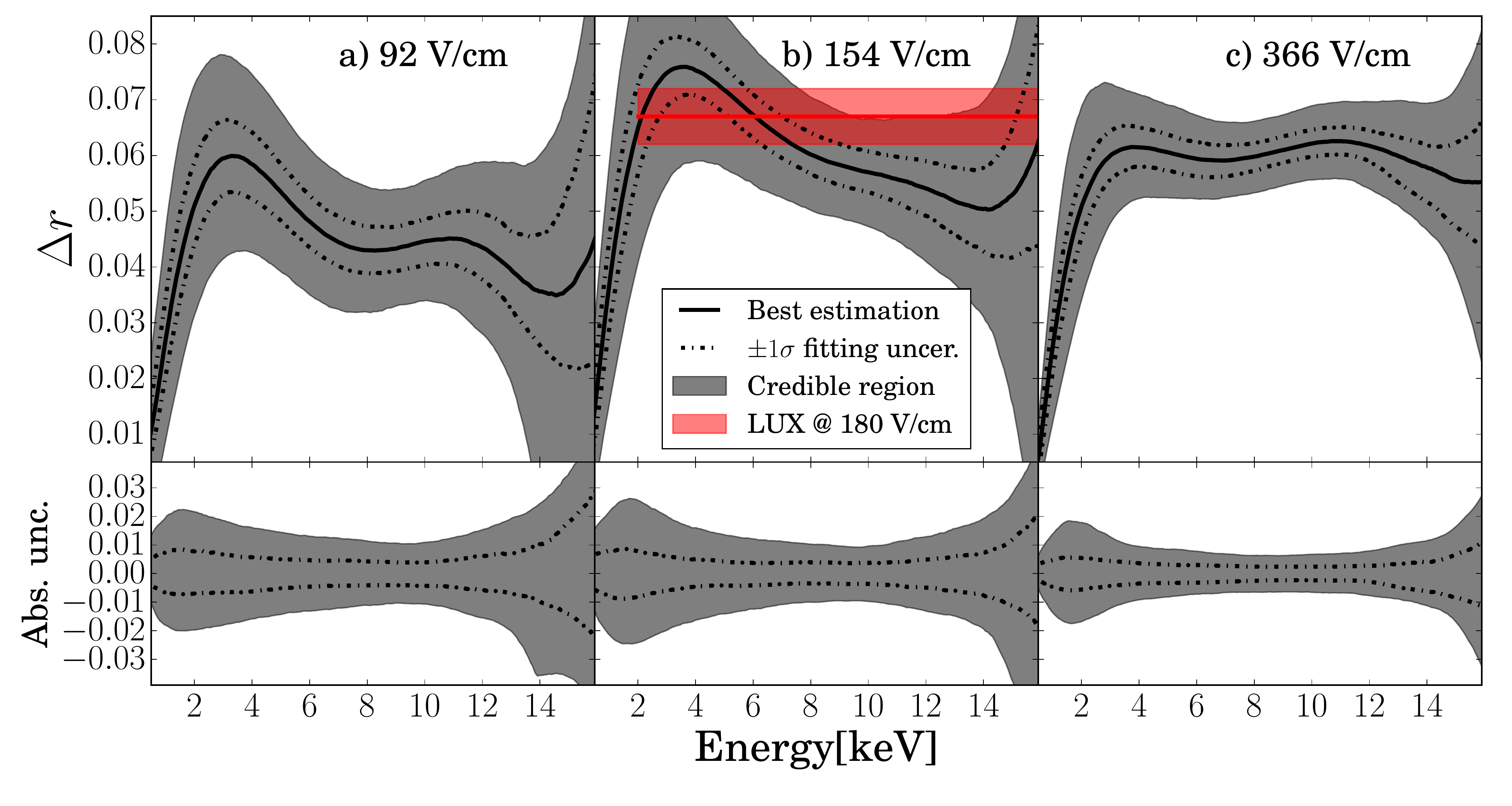}
\caption{
The best estimate for the recombination fluctuation $\Delta r$ as a function of deposited energy.
The panels a), b) and c) show the recombination fluctuations for the three drift fields scanned in the study.
The solid lines represent the mean values and the shaded regions indicate the 15.4\% to 84.6\% credible regions of $\Delta r$.
The dot-dashed lines indicate the fitting uncertainties.
Measured $\Delta r$ values at 180\,V/cm from LUX are shown in panel b) in red solid (mean) and shaded region (uncertainty).
}
\label{fig:DeltaR}
\end{figure*}

The most relevant parameters in this work are the mean recombination fraction $\langle r \rangle$ and the recombination fluctuation $\Delta r$ defined in Eq.~\eqref{eq:RecombinationAndFluctuations1}, respectively. The mean recombination fraction affects the ratio between $n_e$ and $n_{ph}$, thus the band mean in the Log$_{10}(S2/S1)$ vs. $S1$ distribution, while $\Delta r$ affects the variance of the distribution.
In the fit, both $\langle r \rangle$ and $\Delta r$ are parameterized as 4th order polynomial function of the energy deposition $E$ with respect to reference curves. 
The reference curves for $\langle r \rangle$ and $\Delta r$ are initially chosen from NEST v0.98~\cite{NEST:v0.98} and the LUX measurements~\cite{Akerib:2015wdi}, respectively.
The fit results for $\langle r \rangle$, along with the derived mean photon yields $\langle n_{ph} \rangle /E$ and charge yields $\langle n_e \rangle / E$, and $\Delta r$ are shown in Fig.~\ref{fig:PYandCYandR} and~\ref{fig:DeltaR}, respectively. The mean photon and electron yields, $\langle n_{ph} \rangle$ and $\langle n_{e} \rangle$ per unit energy, are calculated via:

\begin{equation}
\label{eq:MeanPhotonChargeYields}
\begin{split}
\frac{\langle n_{ph} \rangle}{E} = \frac{1}{W}\frac{\langle r \rangle+\alpha}{1+\alpha} \\  
\frac{\langle n_{e} \rangle}{E} =  \frac{1}{W}\frac{1-\langle r \rangle}{1+\alpha}.
\end{split}
\end{equation}

The best estimations of $\langle n_{ph} \rangle / E $, $\langle n_e \rangle / E$, $\langle r \rangle$, and $ \Delta r $ are evaluated as the weighted averages of the point estimations over all FVs.
The credible regions of these averages, shown in Fig.~\ref{fig:PYandCYandR} and~\ref{fig:DeltaR} as dashed lines, address the fitting uncertainties. These include both statistical uncertainties and uncertainties from the nuisance parameters priors, such as the exciton-to-ion ratio $n_{ex}/n_i$ which is estimated to be $0.15_{-0.05}^{+0.04}$ from the posteriors of the fittings. 
The credible regions, which include both the systematic and fitting uncertainties shown in Fig.~\ref{fig:PYandCYandR} and~\ref{fig:DeltaR} as the shaded regions, are evaluated based on the equally weighted combination of the posteriors in each FV.


The photon yields obtained from our data are consistent with results reported by LUX~\cite{Akerib:2015wdi} at the two lower fields. The curves from NEST v0.98~\cite{NEST:v0.98} are plotted for comparison, showing a larger deviation especially at higher energy, especially for the two larger fields. 
Above 14\,keV the dominant uncertainties are from the fit due to the small statistics near the endpoint energy of the tritium beta decay.
The increased uncertainties below 2\,keV are due to the $S1$ detection efficiency drop below 5\,PE.
In most of the energy region, the systematic uncertainties, which include the uncertainties from position reconstruction and drift field non-uniformity, are compatible with the statistical uncertainties.

Because the recombination fluctuation affects the tail of the ER distribution significantly and with fewer statistics in the tail region we get larger statistical fluctuations for $\Delta r$. Thus the relative uncertainties for $\Delta r$ are larger than the ones for $\langle n_{ph} \rangle / E$.

\section{\label{sec:disc}Electronic and Nuclear Recoils Discrimination}

The different response of electronic and nuclear recoils in liquid xenon provides a powerful method to reject the dominant electronic recoil background from radioactive materials surrounding the target, decays of internal radioactive contaminants, such as $^{85}$Kr and $^{222}$Rn, and eventually the electron scattering from solar neutrinos~\cite{Baudis:2014neutrino}, as well as the signal fluctuations, which include the recombination fluctuations $ \Delta r $, the instrumental and the statistical fluctuations. A larger difference of the ER/NR recombination factors and smaller $\Delta r$ and statistical fluctuations  will lead to a better ER rejection power. Since the electron-ion recombination factor for electronic recoils is more significantly affected by the electric field than nuclear recoils are, the ER and NR band separation is greater at larger drift fields.  However, a larger drift field will suppress the primary scintillation light, leading to a smaller prompt signal and thus larger statistical fluctuations. The interplay between these factors affects the overall ER rejection power. Previous experiments~\cite{Angle:2007uj,Alner:2007ja,Lebedenko:2008gb,Aprile:2012nq,Akerib:2013tjd,Tan:2016zwf} reported ER rejection powers between 99\% to 99.99\% at about 50\% NR acceptance at different drift fields. The photon detection efficiencies from these experiments are also different. 

\begin{figure}[htbp]
\begin{center}
\includegraphics[width=1.05\columnwidth]{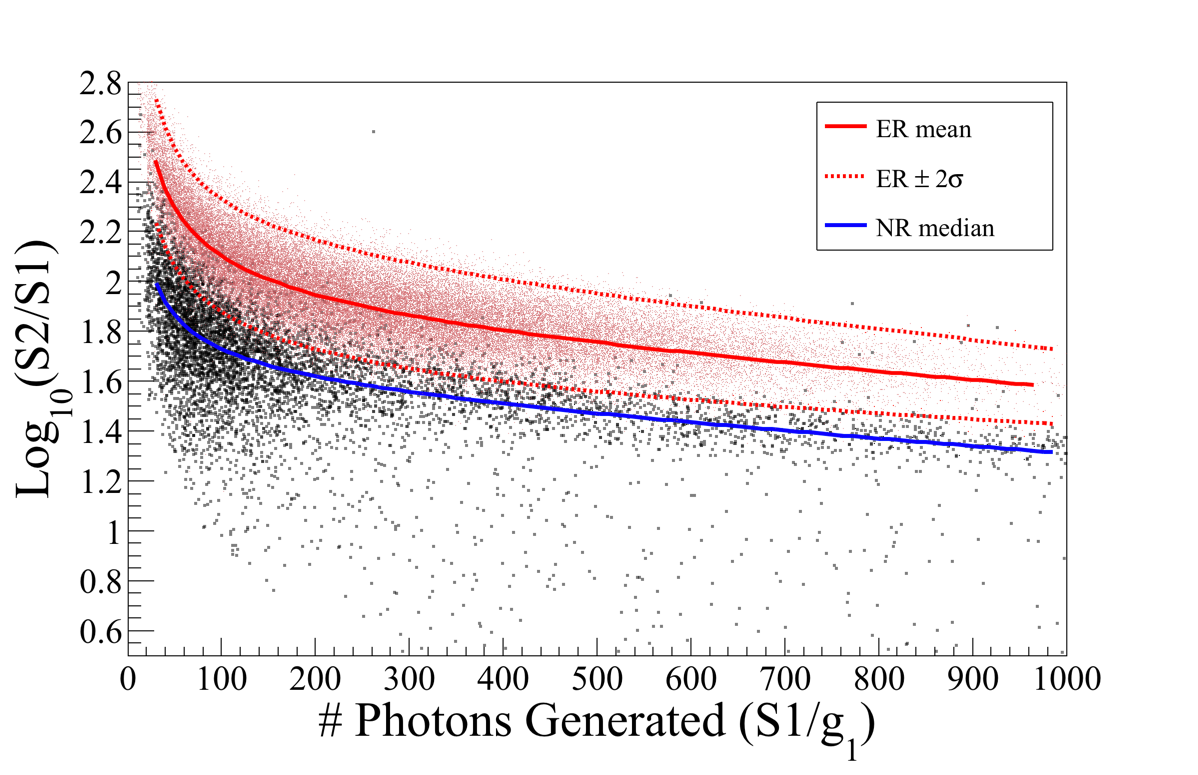}
\includegraphics[width=1.05\columnwidth]{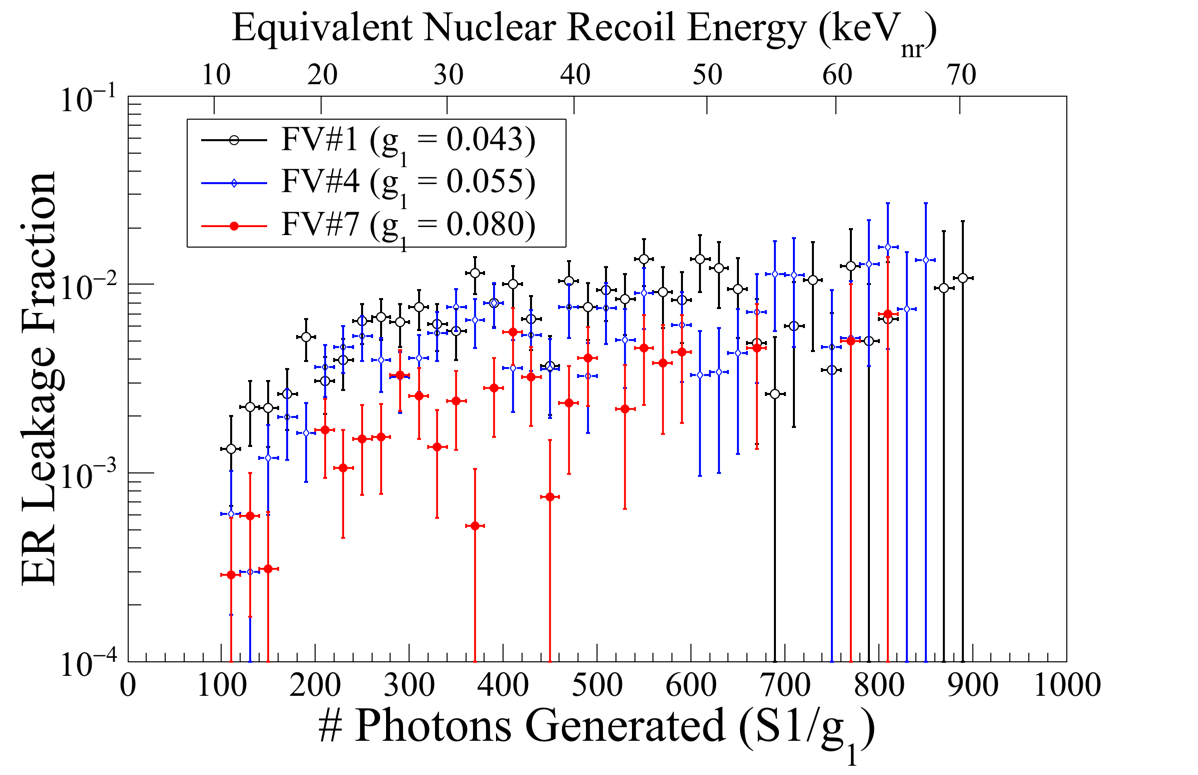}
\caption{(Top) ER (red) and NR (gray) bands from the CH$_3$T and $^{241}$AmBe data at 366~V/cm in FV\#7, which has the largest $g_1$ value among the seven FVs shown in Fig~\ref{fig:fv}. The $S2$ signal is corrected for the electron lifetime. The mean and $\pm2\sigma$ values of the ER band and the median of the NR band are fit by a power law plus a first order polynomial. (Bottom) The ER leakage fractions obtained by counting the number of events below the NR median, divided by the total number of ER events in each bin, for three different FVs from the top to the bottom of the detector at 366~V/cm drift field. The equivalent nuclear recoil energy is calculated based on the $S1$ signal following the method in~\cite{Aprile:2016swn}. 
}
\label{fig:band}
\end{center}
\end{figure}

\begin{figure}[htbp]
\begin{center}
\includegraphics[width=1.0\columnwidth]{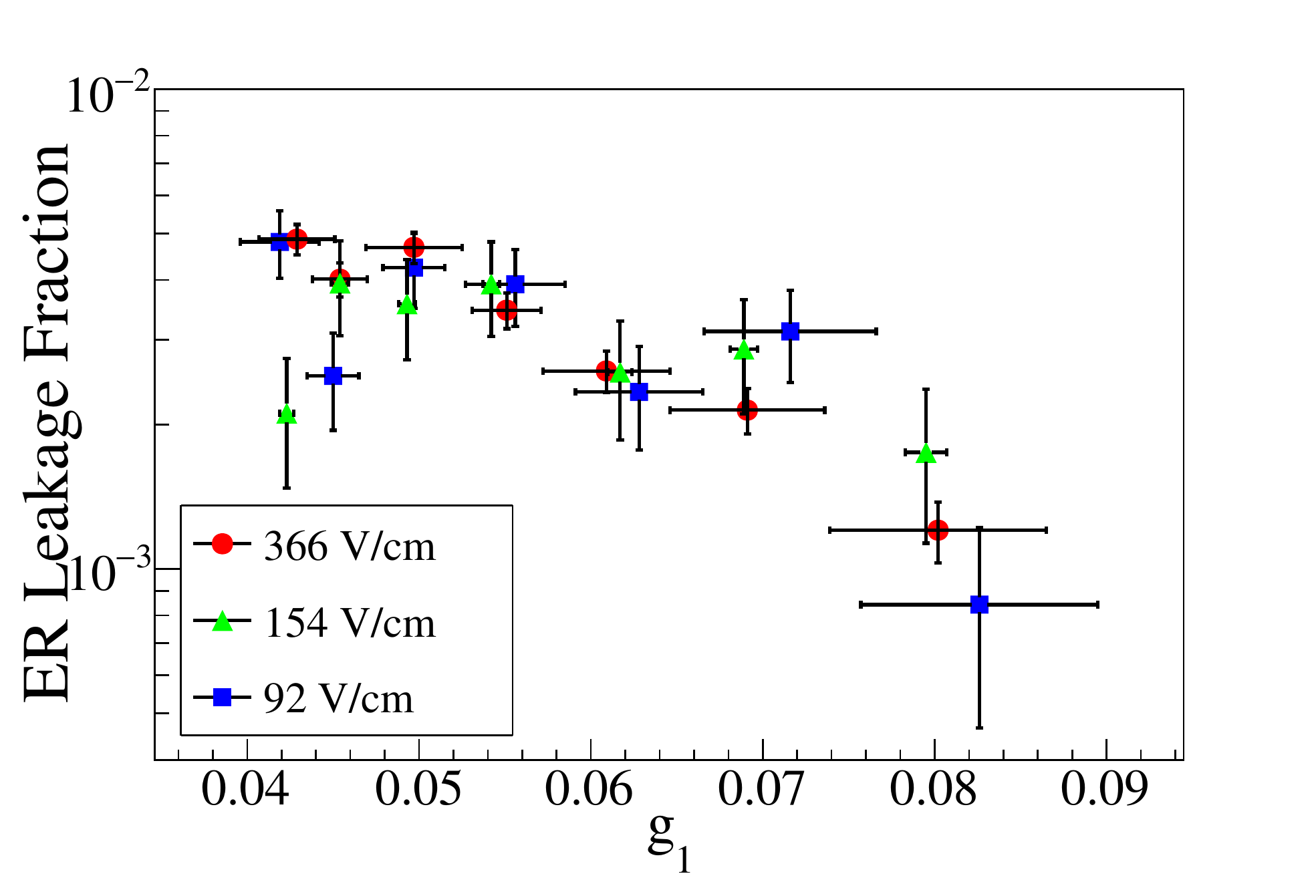}
\caption{The total ER leakage below the nuclear recoil median for different photon detection efficiencies at three measured drift fields. We chose events with S1 corresponding to between 100 and 400 photons generated in liquid xenon. Such a range gives nuclear recoil equivalent energies between approximately 11 and 34 $\rm keV_{nr}$. A smaller ER leakage fraction is observed at a higher photon detection efficiency. The impact of the drift fields on the ER rejection is insignificant for the three fields reported here.   
}
\label{fig:rej}
\end{center}
\end{figure}

Here we use our data to investigate the impact of drift field and photon detection efficiency on the ER rejection power. Fig.~\ref{fig:band} (top) shows an example of the ER and NR bands in the parameter space of Log$_{10}(S2/S1)$ vs. $n_{ph}=S1/g_1$ from the CH$_3$T and $^{241}$AmBe data at 366~V/cm for FV\#7. Normalizing the $S1$ to the number of generated photons, $S1/g_1$, allows us to compare the ER/NR discrimination at the same energy for fiducial volumes with different $g_1$ values. The ER leakage is smaller at lower energies due to the larger separation between the ER and NR bands. Less than 10$^{-3}$ ER leakage is achieved between (10-20)\,keV$_{nr}$ for FV\#7 with about 8\% photon detection efficiency. As expected, for FVs with lower photon detection efficiencies, the ER leakage fraction increases as shown in Fig.~\ref{fig:band} (bottom). This is caused by the larger statistical fluctuations introduced by lower light yields.  

To compare the ER leakage at different drift fields from 92~V/cm to 366~V/cm, we chose an S1 range between 100-400 primary scintillation photons generated in liquid xenon. This corresponds to a NR equivalent energy range of approximately 11-34\,keV$_{nr}$. The dependence of the ER leakage on different photon detection efficiencies is shown in Fig.~\ref{fig:rej} for the seven FVs at the three drift fields studied. The ER rejection power (1 - ER leakage fraction) improves at a higher photon detection efficiency, reaching 99.9\% for $g_1\approx0.08$ at the lowest studied energy of around 10 keV$_{nr}$. 
We did not observe any significant difference for the ER rejection power between the 92\,V/cm and 366\,V/cm drift fields, which is consistent with observations from other dark matter detectors LUX~\cite{Akerib:2016vxi}, PandaX-II~\cite{Tan:2016zwf} and XENON1T~\cite{Aprile:2017iyp}. Although the ER/NR band separation increases from 92\,V/cm to 366\,V/cm, the ER band width (fluctuation) increases, countering the effect on the ER rejection power.

We note that the ER/NR discrimination study presented here is for two specific calibration sources: ER from tritium beta decays and NR from $^{241}$AmBe neutrons. Although it gives a good comparison at different drift fields and photon detection efficiencies, the true ER leakage fraction in a dark matter detector will depend on the background source spectrum, which is different from the tritium beta spectrum with an end-point at $18.6\,$keV. We observed that the ER band width from the tritium beta decays is narrower than that from Compton scatters of external gamma rays and from the background in XENON100. The detailed study of ER band widths and comparison between tritium and other sources can be found in~\cite{Constanze:2017}.

\section{Conclusion}
\label{sec:conclusion}
We report results on the measurement of photon yields and recombination fluctuations for low-energy electronic recoils from tritium beta decays in the XENON100 dark matter detector at three different drift fields (92\,V/cm, 154\,V/cm and 366\,V/cm). We found consistent values compared to those measured by LUX~\cite{Akerib:2015wdi}. By comparing the response between electronic and nuclear recoils at different drift fields and at small fiducial volumes with different photon detection efficiencies, we didn't observe any significant field-dependence of the ER/NR discrimination power between 92\,V/cm and 366\,V/cm. An improvement of the ER rejection power at higher photon detection efficiencies is observed, especially in the low-energy region of interest for dark matter searches. The results provide new information that is relevant to the design, operation and calibration of current and future liquid xenon-based dark matter detectors~\cite{Aprile:2015uzo,Liu:2017nphy,Mount:2017qzi,Aalbers:2016jon}.



\section*{Acknowledgement}
We gratefully acknowledge support from the National Science Foundation, Swiss National Science Foundation, Deutsche Forschungsgemeinschaft, Max Planck Gesellschaft, German Ministry for Education and Research, Netherlands Organisation for Scientific Research, Weizmann Institute of Science, I-CORE, Initial Training Network Invisibles (Marie Curie Actions, PITNGA-2011-289442), Fundacao para a Ciencia e a Tecnologia, Region des Pays de la Loire, Knut and Alice Wallenberg Foundation, Kavli Foundation, and Istituto Nazionale di Fisica Nucleare. We are grateful to Laboratori Nazionali del Gran Sasso for hosting and supporting the XENON project.

\section*{Bibliography}
\bibliographystyle{h-physrev}
\bibliography{bibliography}

\begin{thebibliography}{10}

\bibitem{Chattopadhyay:2003xi}
U.~Chattopadhyay, A.~Corsetti, and P.~Nath,
\newblock Phys. Rev. {\bf D68}, 035005 (2003), hep-ph/0303201.

\bibitem{Akerib:2016vxi}
LUX, D.~S. Akerib {\em et~al.},
\newblock Phys. Rev. Lett. {\bf 118}, 021303 (2017), 1608.07648.

\bibitem{Tan:2016zwf}
PandaX-II, A.~Tan {\em et~al.},
\newblock Phys. Rev. Lett. {\bf 117}, 121303 (2016), 1607.07400.

\bibitem{Aprile:2017iyp}
XENON, E.~Aprile {\em et~al.},
\newblock (2017), 1705.06655.

\bibitem{Cui:2017nnn}
PandaX-II, X.~Cui {\em et~al.},
\newblock (2017), 1708.06917.

\bibitem{Aprile:2015uzo}
XENON, E.~Aprile {\em et~al.},
\newblock JCAP {\bf 1604}, 027 (2016), 1512.07501.

\bibitem{Liu:2017nphy}
J.~Liu, X.~Chen, and X.~Ji,
\newblock Nat Phys {\bf 13}, 212 (2017).

\bibitem{Mount:2017qzi}
B.~J. Mount {\em et~al.},
\newblock (2017), 1703.09144.

\bibitem{Aalbers:2016jon}
DARWIN, J.~Aalbers {\em et~al.},
\newblock JCAP {\bf 1611}, 017 (2016), 1606.07001.

\bibitem{Sikivie:1983ip}
P.~Sikivie,
\newblock Phys. Rev. Lett. {\bf 51}, 1415 (1983),
\newblock [Erratum: Phys. Rev. Lett.52,695(1984)].

\bibitem{Redondo:2013wwa}
J.~Redondo,
\newblock JCAP {\bf 1312}, 008 (2013), 1310.0823.

\bibitem{Aprile:2015ade}
XENON100, E.~Aprile {\em et~al.},
\newblock Science {\bf 349}, 851 (2015), 1507.07747.

\bibitem{Manzur:2009hp}
A.~Manzur {\em et~al.},
\newblock Phys.Rev. {\bf C81}, 025808 (2010), 0909.1063.

\bibitem{Plante:2011hw}
G.~Plante {\em et~al.},
\newblock Phys.Rev. {\bf C84}, 045805 (2011), 1104.2587.

\bibitem{Aprile:2013teh}
XENON100, E.~Aprile {\em et~al.},
\newblock Phys. Rev. {\bf D88}, 012006 (2013), 1304.1427.

\bibitem{Akerib:2016mzi}
LUX, D.~S. Akerib {\em et~al.},
\newblock (2016), 1608.05381.

\bibitem{Lenardo:2014cva}
B.~Lenardo {\em et~al.},
\newblock IEEE Trans. Nucl. Sci. {\bf 62}, 3387 (2015), 1412.4417.

\bibitem{Aprile:2012an}
E.~Aprile {\em et~al.},
\newblock Phys. Rev. {\bf D86}, 112004 (2012), 1209.3658.

\bibitem{Baudis:2013cca}
L.~Baudis {\em et~al.},
\newblock Phys. Rev. {\bf D87}, 115015 (2013), 1303.6891.

\bibitem{Akimov:2014cha}
D.~{\relax Yu}. Akimov {\em et~al.},
\newblock JINST {\bf 9}, P11014 (2014), 1408.1823.

\bibitem{Lin:2015jta}
Q.~Lin {\em et~al.},
\newblock Phys. Rev. {\bf D92}, 032005 (2015), 1505.00517.

\bibitem{Goetzke:2016lfg}
L.~W. Goetzke, E.~Aprile, M.~Anthony, G.~Plante, and M.~Weber,
\newblock (2016), 1611.10322.

\bibitem{Akerib:2016qlr}
LUX, D.~S. Akerib {\em et~al.},
\newblock Phys. Rev. {\bf D95}, 012008 (2017), 1610.02076.

\bibitem{NEST:v0.98}
M.~Szydagis {\em et~al.},
\newblock JINST {\bf 6}, P10002 (2011), 1106.1613.

\bibitem{Aprile:2016swn}
XENON100, E.~Aprile {\em et~al.},
\newblock Phys. Rev. {\bf D94}, 122001 (2016), 1609.06154.

\bibitem{Aprile:2011dd}
XENON100, E.~Aprile {\em et~al.},
\newblock Astropart.Phys. {\bf 35}, 573 (2012), 1107.2155.

\bibitem{Aprile:2016pmc}
XENON, E.~Aprile {\em et~al.},
\newblock Phys. Rev. {\bf D95}, 072008 (2017), 1611.03585.

\bibitem{Akerib:2015wdi}
LUX, D.~S. Akerib {\em et~al.},
\newblock Phys. Rev. {\bf D93}, 072009 (2016), 1512.03133.

\bibitem{Dahl:2009nta}
C.~E. Dahl,
\newblock {\em {The physics of background discrimination in liquid xenon, and
  first results from Xenon10 in the hunt for WIMP dark matter}},
\newblock PhD thesis, Princeton U., 2009.

\bibitem{Baudis:2010Kr83m}
A.~Manalaysay {\em et~al.},
\newblock Rev. Sci. Instrum. {\bf 81}, 073303 (2010), 0908.0616v2.

\bibitem{Fano_Process}
U.~Fano,
\newblock Phys. Rev. {\bf 72}, 26 (1947).

\bibitem{Doke_Fano_Estimate}
A.~Doke, T. an~Hitachi,
\newblock Nucl. Instrum. Meth. {\bf A134}, 353 (1976).

\bibitem{GelmanMCMCGoF}
A.~Gelman, X.~Meng, and H.~Stern,
\newblock Statistica Sinica {\bf 6}, 733 (1996).

\bibitem{DPEMeasurement}
C.~Faham {\em et~al.},
\newblock JINST {\bf 10}, P09010 (2015), 1506.08748.

\bibitem{Lansiart:1976}
A.~Lansiart, A.~Seigneur, J.-L. Moretti, and J.-P. Morucci,
\newblock Nucl. Instrum. A {\bf 135}, 47 (1976).

\bibitem{Aprile:2013blg}
XENON100, E.~Aprile {\em et~al.},
\newblock J. Phys. {\bf G41}, 035201 (2014), 1311.1088.

\bibitem{BetaSpecKatrin}
G.~Drexlin, V.~Hannen, S.~Mertens, and C.~Weinheimer,
\newblock Adv. High Energy Phys. {\bf 2013}, 293986 (2013).

\bibitem{AIMCMC}
J.~Goodman and J.~Weare,
\newblock CAMCoS {\bf 5}, 65 (2010).

\bibitem{EMCEE}
D.~Foreman-Mackey, W.~Hogg, D.~Lang, and J.~Goodman,
\newblock Publ. Astron. Soc. Pac. {\bf 125}, 925 (2013), 1202.3665.

\bibitem{Baudis:2014neutrino}
L.~Baudis {\em et~al.},
\newblock JCAP {\bf 01}, 044 (2014), 1309.7024.

\bibitem{Angle:2007uj}
XENON, J.~Angle {\em et~al.},
\newblock Phys.Rev.Lett. {\bf 100}, 021303 (2008), 0706.0039.

\bibitem{Alner:2007ja}
G.~J. Alner {\em et~al.},
\newblock Astropart. Phys. {\bf 28}, 287 (2007), astro-ph/0701858.

\bibitem{Lebedenko:2008gb}
V.~N. Lebedenko {\em et~al.},
\newblock Phys. Rev. {\bf D80}, 052010 (2009), 0812.1150.

\bibitem{Aprile:2012nq}
XENON100, E.~Aprile {\em et~al.},
\newblock Phys. Rev. Lett. {\bf 109}, 181301 (2012), 1207.5988.

\bibitem{Akerib:2013tjd}
LUX, D.~S. Akerib {\em et~al.},
\newblock Phys. Rev. Lett. {\bf 112}, 091303 (2014), 1310.8214.

\bibitem{Constanze:2017}
C.~Hasterok,
\newblock {\em Gas Purity Analytics, Calibration Studies, and Background
  Predictions, towards the First Results of XENON1T},
\newblock PhD thesis, 2017.

\end{thebibliography}

\end{document}